\documentclass[conference]{IEEEtran}
\IEEEoverridecommandlockouts
\usepackage{cite}
\usepackage{amsmath,amssymb,amsfonts}
\usepackage{algorithmic}
\usepackage{graphicx}
\usepackage{textcomp}
\usepackage{xcolor}
\usepackage{algorithm}
\usepackage{array}
\usepackage[caption=false,font=normalsize,labelfont=scriptsize,textfont=scriptsize]{subfig}
\usepackage{stfloats}
\usepackage{url}
\usepackage{verbatim}
\usepackage{subfig}
\usepackage{amssymb}
\usepackage{siunitx}
\usepackage{threeparttable}
\usepackage{bm}
\usepackage{authblk}
\allowdisplaybreaks[3]
\def\BibTeX{{\rm B\kern-.05em{\sc i\kern-.025em b}\kern-.08em
    T\kern-.1667em\lower.7ex\hbox{E}\kern-.125emX}}
\begin{document}

\title{Environment-Aware Codebook for RIS-Assisted MU-MISO Communications: \\Implementation and Performance Analysis} 
\author[1,2]{Zhiheng Yu}
\author[3,$\dag$]{Jiancheng An}
\author[1,2]{Lu Gan}
\author[3]{Chau Yuen \thanks{This work was supported in part by the Ministry of Education (MOE) Tier 2 (Award Number MOE-T2EP50220-0019).}}
\affil[1]{School of Information and Communication Engineering, \authorcr University of Electronic Science and Technology of China (UESTC), Chengdu, 611731, China.}
\affil[2]{Yibin Institute of UESTC, Yibin 644000, China.}
\affil[3]{School of Electrical and Electronics Engineering, Nanyang Technological University, Singapore 639798. \authorcr Email: jiancheng\underline{~}an@163.com $^{\dag}$ Corresponding author: Jiancheng An}

\renewcommand*{\Affilfont}{\normalsize} 
\renewcommand\Authands{ and } 
\date{} 

\maketitle

\begin{abstract}
Reconfigurable intelligent surface (RIS) provides a new electromagnetic response control solution, which can reshape the characteristics of wireless channels. In this paper, we propose a novel environment-aware codebook protocol for RIS-assisted multi-user multiple-input single-output (MU-MISO) systems. Specifically, we first introduce a channel training protocol which consists of off-line and on-line stages. Secondly, we propose an environment-aware codebook generation scheme, which utilizes the statistical channel state information and alternating optimization method to generate codewords offline. Then, in the on-line stage, we use these pre-designed codewords to configure the RIS, and the optimal codeword resulting in the highest sum rate is adopted for assisting in the downlink data transmission. Thirdly, we analyze the theoretical performance of the proposed protocol considering the channel estimation errors. Finally, numerical simulations are provided to verify our theoretical analysis and the performance of the proposed scheme.
\end{abstract}

\section{Introduction}
\IEEEPARstart{R}{econfigurable} intelligent surface (RIS) is an artificial electromagnetic (EM) surface with programmable EM properties \cite{r1}. In contrast to conventional wireless technologies like millimeter wave (mmWave) communication that have to grapple with issues related to network energy consumption and hardware costs, the RIS elements are passive and simply reflect signals, mitigating the power-thirsty radio frequency chains and forward delay in traditional relays \cite{JSAC_2023_An_Stacked, cao2021ai,CL_2023_An_A1, kan2023hybrid}. Remarkably, RIS can operate in full-duplex mode without self-interference problems \cite{arXiv_2023_An_Stacked, TWC_2023_Xu_Antenna}. Various functionalities like indoor positioning can be achieved by adjusting the RIS reflection coefficient (RC)\cite{VTC1}. Given these advantages, RIS is considered a key technology enabling future wireless networks.

Nevertheless, RIS-assisted wireless communication systems also present some challenges, one of which is the optimization of RC at the RIS\cite{TVT_2023_Xu_Channel, r19,r13,liu2024drl, VTC2,r15,add9,add25}. Specifically, in \cite{r19}, \emph{Wu and Zhang} minimized the total transmit power under specific signal-to-interference-plus-noise ratio (SINR) constraints, they adopted alternating optimization (AO) based on semidefinite relaxation (SDR) to obtain approximate solutions of transmit precoding and RIS RCs. In \cite{r13}, \emph{Yan et al.} devised RCs based on the statistical channel state information (CSI) to maximize spectral efficiency. In \cite{VTC2}, \emph{Ma et al.} developed a deep deterministic policy gradient (DDPG) algorithm to solve the joint optimization problem of beamforming at the base station (BS) and RCs at the RIS. However, the schemes above all assume the perfect CSI, which is generally unavailable in practical scenarios. Considering the imperfect CSI, \emph{An et al.} conceived a low-complexity framework for maximizing the achievable rate of RIS-aided multiple-input multiple-output (MIMO) systems with discrete phase shifts \cite{r15}. Moreover, \cite{add9} considered an imperfect CSI scenario and solved the max-min SINR problem with continuous phase shifts. In order to reduce the pilot overhead, \emph{Shen et al.} proposed a novel RIS architecture based on group and fully connected reconfigurable impedance networks and derived the scaling law for received signal power in RIS-aided MIMO systems \cite{add25}. 

Recently, various codebook-based schemes have emerged to enhance RIS performance within moderate overhead. For instance, \cite{add21} proposed a codebook-based framework and introduced three codebook generation schemes. Furthermore, \cite{r9} designed the random and uniform codebooks and analyzed their theoretical performance. Besides, \cite{add23} developed a multi-agent deep reinforcement learning based beamforming codebook design that relies only on the received power.

Nevertheless, existing codebook-based schemes usually entail substantial training overhead and lack adaptability to channel environment variations. Against this background, in this paper, we introduce a codebook-based protocol and propose an environment-aware scheme to generate the discrete RIS RC configuration codebook. The proposed scheme has the adaptability to the channel environment and can strike a beneficial trade-off between training overhead and system performance. In addition, we theoretically analyze the received power scaling law in a single user scenario considering both perfect and imperfect CSI.

\section{System Model}

\begin{figure}[t]
\centering
\includegraphics[width=8.8cm]{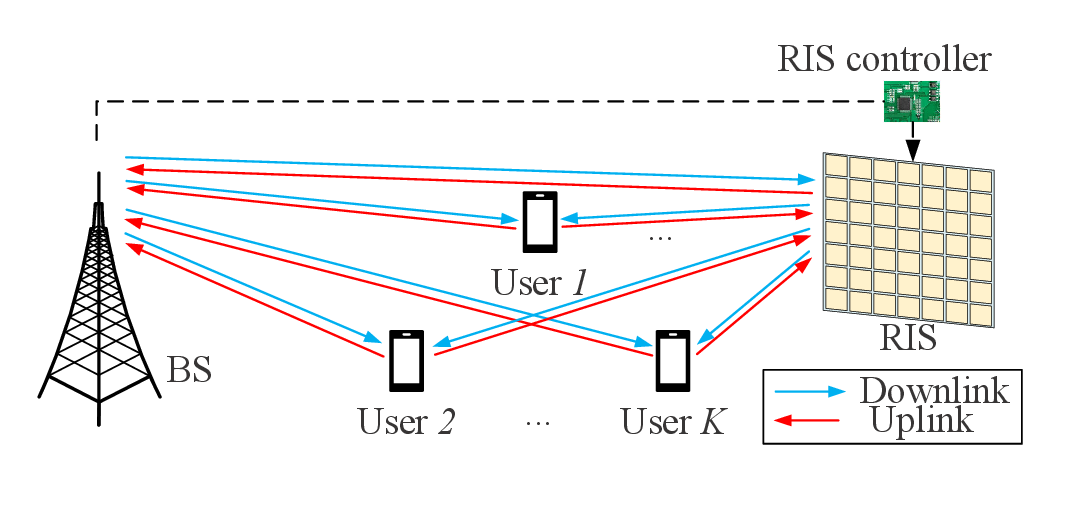}
\caption{A RIS-assisted multi-user communication system.}
\label{fig_1}
\end{figure}

Let us consider a RIS-assisted multi-user communication system in a single cell as shown in Fig. \ref{fig_1}, where a RIS with $N$ reflecting elements is deployed to enhance the downlink (DL) signal transmission between a multi-antenna BS and $K$ single-antenna users. The numbers of antennas at the BS and RIS elements are denoted by $M$ and $N$, respectively. The RIS is equipped with a smart controller capable of adjusting the RCs according to the real-time CSI. In this paper, we adopt a quasi-static flat-fading channel model for all links. Besides, we consider a time-division duplexing (TDD) protocol for both uplink (UL) as well as DL transmissions and assume the channel’s reciprocity for the CSI acquisition in the DL based on the UL training. The baseband equivalent channels spanning from the BS to the RIS, from the RIS to user $k$, and from the BS to user $k$ are denoted by $\mathbf{G} \in \mathbb{C}^{N \times M}$, $\mathbf{h}_{r,k}^H \in \mathbb{C}^{1 \times N}$, $\mathbf{h}_{d,k}^H \in \mathbb{C}^{1 \times M}$, respectively, with $k = 1, 2, \cdots, K$.

Let the diagonal matrix $\mathbf{\Phi} = \text{diag} \left(\varphi_1, \varphi_2, \cdots, \varphi_N \right)$ represent the RC configuration of the RIS, where $\varphi_n$ denotes the RC of the $n$th RIS element, following $\left| \varphi_n \right| = 1$ for $n = 1,2,\cdots,N$. In this paper, we use discrete phase shifts to control signal reflection. In order to facilitate the practical implementation, we consider that the discrete phase shifts are obtained by uniformly quantizing the interval $\left[0, 2\pi \right)$ and let $b$ represent the number of bits used for quantizing the phase shift levels. Hence the phase shift of each RIS element is assumed to value from discrete phase shift set $\mathcal B = \left \{0, \varDelta \theta, \cdots, (2^b-1) \varDelta \theta \right \}, $ where $\varDelta \theta =2 \pi /2^b $. Thus, the RC of the $n$th element at the RIS can be expressed as $\varphi_n = e^{j \theta_n} $, $\theta_n \in \mathcal{B}$, $n = 1,2,\cdots,N$. Therefore, the composite end-to-end channel for user $k$ is given by
\begin{align} \mathbf{h}_k^H = \mathbf{h}_{r,k}^H \mathbf{\Phi} \mathbf{G} + \mathbf{h}_{d,k}^H, \quad k = 1, 2, \cdots, K. \end{align}  
Let $\mathbf{H} = \left[ \mathbf{h}_1, \mathbf{h}_2, \cdots, \mathbf{h}_K \right] \in \mathbb{C}^{M \times K}$ represent the UL channels associated with $K$ users. During the channel estimation process, the pilot signal received at the BS is given by \setlength{\abovedisplayskip}{2pt} \setlength{\belowdisplayskip}{2pt}
\begin{align} \mathbf{y} = \sum_{k=1}^{K} \mathbf{h}_k \sqrt{P_{\text{ul}}} x_k + \mathbf{z}, \end{align} 
where $P_{\text{ul}}$ is the average power of the pilot symbols, which is assumed to be identical for all users, $x_k$ denotes the pilot symbol transmitted from user $k$ to BS with zero mean and unit variance, $\mathbf{z} \in \mathbb{C}^{M \times 1}$ denotes the additive white Gaussian noise (AWGN) at the BS obeying $\mathbf{z} \sim \mathcal{CN} \left( \mathbf{0},\sigma_z^2 \mathbf{I}_M \right)$, $\mathbf{I}_M \in \mathbb{C}^{M \times M}$ denotes the identity matrix. 

Next, we consider the DL of data transmission, and use linear transmit precoding at the BS. As a result, the complex baseband signal received at user $k$ is given by 
\begin{align} r_k= \mathbf{h}_k^H \sum_{i=1}^{K} \mathbf{w}_i s_i + n_k, \end{align}
where $s_i$ denotes the transmitted information symbol of user $i$, $\mathbf{w}_i \in \mathbb{C}^{M \times 1}$ denotes the corresponding transmit beamforming vector and $n_k \sim \mathcal{CN} \left( 0,\sigma_k^2 \right), k = 1,2,\cdots,K $ denotes the AWGN at the $k$th user’s receiver.

Meanwhile, we adopt the Rician channel in this paper \cite{jia2023codebook}. Specifically, the RIS-user $k$ channel can be expressed as 
\begin{align} \label{eq:5} \mathbf{h}_{r,k} = \sqrt{\beta_r} \left( \sqrt{\frac{F_r}{F_r + 1}}\mathbf{h}_{r,k}^{\text{LoS}} + \sqrt{\frac{1}{F_r + 1}}\mathbf{h}_{r,k}^{\text{NLoS}} \right), \end{align} 
where $\beta_r$ and $F_r$ are the path loss and the Rician factor of the RIS-user channel, respectively; $\mathbf{h}_{r,k}^{\text{LoS}}$ and $\mathbf{h}_{r,k}^{\text{NLoS}}$ represent the line-of-sight (LoS) and non-line-of-sight (NLoS) components of $\mathbf{h}_{r,k}$, respectively. The NLoS component is i.i.d. complex Gaussian distributed with zero mean and unit variance, satisfying $\mathbf{h}_{r,k}^{\text{NLoS}} \sim \mathcal{CN} \left(\mathbf{0}, \mathbf{I}_N \right)$. Similarly, the BS-user channel and BS-RIS channel are modeled by using (\ref{eq:5}).

\section{The Proposed Channel Training Protocol}

In this section, we outline the proposed scheme by first introducing our channel training-based protocol in Section III-A. Then, our bespoke environment-aware codebook generation scheme is proposed in Section III-B. In Section III-C, we introduce the on-line codebook configuration stage.

\subsection{Channel Training-Based Protocol}

\begin{figure}[t]
\centering
\includegraphics[width=8.8cm]{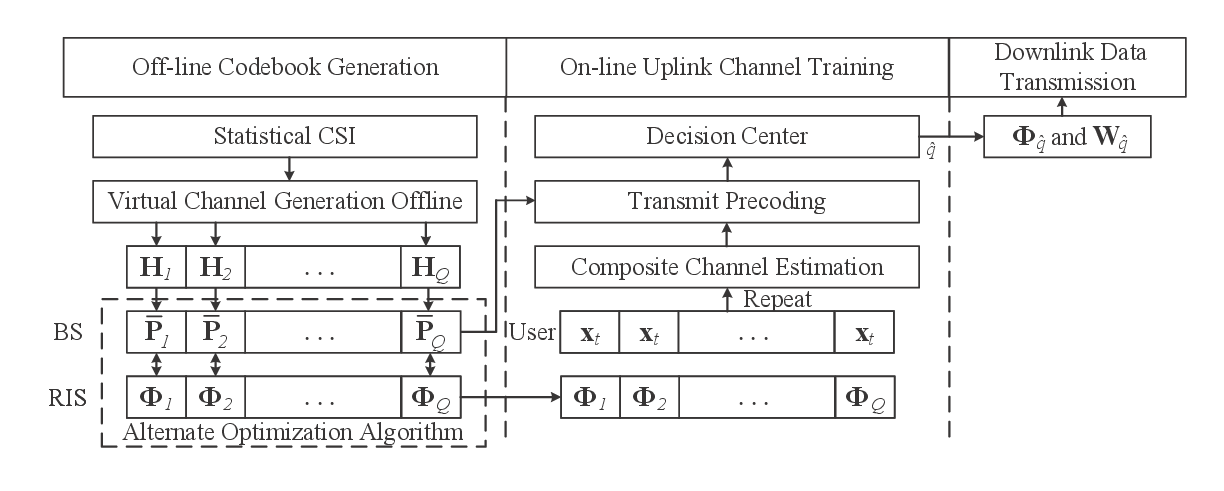}
\caption{The frame structure of the proposed protocol.}
\label{fig_framework}
\end{figure}

As shown in Fig. \ref{fig_framework}, our channel training-based protocol consists of off-line and on-line stages.

In the off-line stage, we generate a group of virtual channels using the statistical CSI and use the AO algorithm to obtain the optimal RC configuration corresponding to each virtual channel. Repeating for $Q$ virtual channels generates the environment-aware codebook.

In the on-line stage, we have two phases in the transmission protocol. Specifically, during the UL channel training phase, we adjust the RIS configuration over training blocks according to the environment-aware codebook. Then, we repeat the composite channel estimation and the transmit precoding to obtain $Q$ candidate channels; During the DL signal transmission phase, we select the optimal channel maximizing the sum rate from $Q$ candidates, and get the corresponding RIS configuration and transmit power allocation to assist in the DL communications. 

\subsection{Off-Line Codebook Design Stage}

\subsubsection{Virtual Channel Generation Based on Statistical CSI}
Note that the statistical CSI is easier to be obtained and has a longer channel coherence time. Hence according to (\ref{eq:5}), we generate a set of virtual RIS-user $k$ channels as follows \begin{align} \label{eq:12} \mathbf{h}_{r,k,q} = \sqrt{\beta_r} \left( \sqrt{\frac{F_r}{F_r + 1}} \mathbf{h}_{r,k}^{\text{LoS}} + \sqrt{\frac{1}{F_r + 1}} \mathbf{\tilde{h}}_{r,k,q}^{\text{NLoS}} \right), \end{align} where $\mathbf{h}_{r,k}^{\text{LoS}}$ is the LoS component of the virtual channel with unit magnitude and it remains unchanged within each channel coherence block, $\mathbf{\tilde{h}}_{r,k,q}^{\text{NLoS}}$ represents the NLoS component of the $q$th virtual channel which is randomly generated following the same distribution with $\mathbf{h}_{r,k}^{\text{NLoS}}$. In addition, the virtual BS-RIS and BS-user $k$ channels are generated in the same way.

\subsubsection{Alternating Optimization Algorithm}
For each tentative virtual channel, we endeavor to maximize the sum rate for all users by jointly optimizing the transmit precoding at the BS and the RC configuration at the RIS, subject to the total transmit power. Specifically, the optimization problem is formulated as
\begin{align} \label{eq:9} \max_{\mathbf{W}_q, \mathbf{\Phi}_q} & \sum_{k=1}^{K} \text{log}_2 \left( 1 + \frac{\left| \mathbf{h}_{k,q}^H \mathbf{w}_{k,q} \right|^2}{\sum_{l \neq k}^K \left| \mathbf{h}_{k,q}^H \mathbf{w}_{l,q} \right|^2 + \sigma_k^2} \right) \notag\\ \mathrm{s.t.} & \mathbf{\Phi}_q = \text{diag} \left( \varphi_{q,1}, \varphi_{q,2}, \cdots, \varphi_{q,N} \right), \notag\\&  \sum_{k=1}^{K} \lVert \mathbf{w}_{k,q}\rVert^2 = P_d, q = 1, 2, \cdots, Q, \end{align} 
where $\mathbf{h}_{k,q}^H$ and $\mathbf{w}_{k,q}$ are the composite channel and the transmit precoding vector for user $k$ in the $q$th training block, respectively. $\mathbf{W}_q = \left[\mathbf{w}_{1,q}, \mathbf{w}_{2,q}, \cdots, \mathbf{w}_{K,q} \right] \in \mathbb{C}^{M \times K}$ is the transmit precoding matrix, while $P_d$ is the total transmit power at the BS.

We adopt zero-forcing (ZF) beamforming at the BS to eliminate interference between different users and let $p_{k,q}$ denote the received power of user $k$ in the $q$th training block, yielding $ \mathbf{h}_{k,q}^H \mathbf{w}_{k,q} = \sqrt{p_{k,q}}, \mathbf{h}_{k,q}^H \mathbf{w}_{l,q} = 0, \forall l \neq k, k = 1, 2, \cdots, K$.

Let $\mathbf{H}_q^H = \left[\mathbf{h}_{1,q}, \mathbf{h}_{2,q}, \cdots, \mathbf{h}_{K,q} \right]^H \in \mathbb{C}^{K \times M}, q = 1, 2, \cdots, Q$ represent the DL composite channels in the $q$th training block, the optimal transmit precoding matrix $\mathbf{W}_q$ is given by a pseudo-inverse of $\mathbf{H}_q^H$ with an appropriate power allocation among users, which can be expressed as 
\begin{align} \label{eq:w} \mathbf{W}_{q} = \mathbf{H}_{q} \left(\mathbf{H}_{q}^H \mathbf{H}_{q} \right)^{-1} \mathbf{P}_q^{\frac{1}{2}}, \end{align} 
where $\mathbf{P}_q = \text{diag}\left(p_{1,q}, p_{2,q}, \cdots, p_{K,q} \right)$ denotes the received power matrix in the $q$th training block. Consequently, the sum rate can be simplified as $R_q = \sum_{k=1}^{K} \text{log}_2 \left(1 + p_{k,q}/\sigma_k^2 \right)$.

Next, we utilize the AO algorithm to obtain the optimal transmit power allocation and RC solution, given the virtual channel generated using statistical CSI.

\paragraph{Transmit Power Allocation}

Given a tentative RIS RC configuration matrix $\mathbf{\Phi}_q$, $K$ unknown transmit power allocation coefficients can be obtained by applying the classic water-filling power allocation algorithm. Let $\mathbf{U}_q = \left( \mathbf{H}_q^H \mathbf{H}_q \right)^{-1} \in \mathbb{C}^{K \times K}$. Based on (\ref{eq:w}), we can easily deduce that the transmit power allocated to the $k$th user can be expressed as \begin{align} \label{eq:pk} \overline{p}_{k,q} = \text{max} \left\{ \frac{1}{\eta_q} \ \text{---} \ \sigma_k^2 u_{k,q}, 0 \right\}, \end{align} where $u_{k,q}$ is the $k$th diagonal element of $\mathbf{U}_q$ and $\eta_q$ is a normalized factor that satisfies $\sum_{k=1}^{K} \overline{p}_{k,q} = P_d$. Once that the transmit power allocation among users is obtained, the received power of user $k$ is $p_{k,q} = \overline{p}_{k,q}/u_{k,q}.$

\paragraph{RIS RC Optimization}

Given transmit power allocation matrix $\overline{\mathbf{P}}_q$, (\ref{eq:9}) is reduced to
\begin{align} \label{eq:15}  \max_{\theta_{q,n}} &\enspace R_q\left( \theta \right) = \sum_{k=1}^{K} \text{log}_2 \left( 1 + \frac{\overline{p}_{k,q}}{u_{k,q} \sigma_k^2} \right) \notag\\ \mathrm{s.t.} &\enspace \theta_{q,n} \in \mathcal{B}, \forall n \in \mathcal{N}, \end{align} which can be solved by employing the successive refinement method to optimize RIS RCs one-by-one.

We use random phase shifts to initialize power allocation and repeat AO until the objective function $R_q$ reaches convergence. By carrying out the virtual channel generation and the AO method for all $Q$ training blocks, we obtain $\mathbf{\Phi}_q$ and $\overline{\mathbf{P}}_q$, with $q = 1, 2, \cdots, Q$.

\subsection{On-Line RIS Configuration Stage}

In the on-line stage, we complete the RIS configuration by selecting the optimal codeword in the environment-aware codebook to assist in the DL data transmission. 

\subsubsection{Composite Channel Estimation}

Given the RC configuration of the $q$th training block $\mathbf{\Phi}_q$, upon collecting the signals of $T$ time slots (TSs), the signal received at the BS in this block is given by
\begin{align}  \mathbf{Y}_q = \sqrt{P_{\text{ul}}} \mathbf{H}_q \mathbf{X} + \mathbf{Z}_q, \end{align} 
where $\mathbf{X} = \left[ \mathbf{x}_1, \mathbf{x}_2,\cdots, \mathbf{x}_T \right] \in \mathbb{C}^{K \times T}$ is the pilot matrix used for estimating the composite channels of the $K$ users, which is identical for all $Q$ training blocks. $\mathbf{x}_t = [x_{1,t}, x_{2,t}, \cdots, x_{K,t}]^T$ is the pilot vector of the $t$th TS. $\mathbf{Z}_q = \left[ \mathbf{z}_{q,1}, \mathbf{z}_{q,2}, \cdots, \mathbf{z}_{q,T} \right] \in \mathbb{C}^{M \times T}$ denotes the noise matrix at the BS.

The composite channel estimation during each block is equivalent to the channel estimation for conventional multi-user multiple-input single-output (MU-MISO) systems. Adopting the orthogonal pilot design and the minimum pilot overhead $T = K$ \cite{r17}. The least squares (LS) estimate of the composite channel $\mathbf{H}_q$ is given by
\begin{align} \label{eq:8}\widetilde{\mathbf{H}}_{q} = \frac{1}{K \sqrt{P_{\text{ul}}}} \mathbf{Y}_q \mathbf{X}^H.
\end{align}

\subsubsection{Transmit Precoding}

Based on the estimate of the composite channel, BS performs multi-user transmit precoding $\widetilde{\mathbf{W}}_q$ by utilizing the ZF precoder according to (\ref{eq:w}).

\subsubsection{Optimal Index Selection}

Then, after obtaining $Q$ candidate channels and their transmit precoding matrices, the RIS RC configuration problem can be expressed as
\begin{align} \label{eq:11} \max_q & \enspace R_q = \sum_{k=1}^{K} \text{log}_2 \left(1 + \frac{1}{\sigma_k^2} \left|\widetilde{\mathbf{h}}_{k,q}^H \widetilde{\mathbf{w}}_{k,q} \right|^2 \right)  \notag\\ \mathrm{s.t.} & \enspace 1 \leq q \leq Q .\end{align}
Once the index $\hat{q}$ of the optimal training block is obtained, we get the RIS RC configuration $\hat{\mathbf{\Phi}} = \mathbf{\Phi}_{\hat{q}}$, and the corresponding transmit precoding matrix accordingly. Note that different from the traditional AO algorithm, the complexity of the proposed scheme is independent of the RIS elements number $N$, but increases with the increase of the training overhead $Q$.

\section{Theoretical Analysis of the Proposed Scheme}
In this section, we analyze the performance of the proposed environment-aware scheme by considering a single user case. For brevity, we assume that the direct BS-user link is blocked and the BS-RIS channel remains only the LoS component which is likely to happen in practical scenarios. For our channel training protocol, the channel estimation errors $\varepsilon_q \sim \mathcal{CN} \left(0, \sigma_q^2 \right), q = 1, 2, \cdots, Q$ directly influence the composite channels. Hence, the achievable rate maximization problem is equivalent to the problem
\begin{align} \label{eq:p_e} \max_{q \in \mathcal{Q}} \hat{Y}_q = \left| \sum_{n=1}^{N} h_{r,n}^* \varphi_{q,n} g_{n,m} + \varepsilon_q \right|^2.
\end{align}
Note that the influence of channel estimation error is implicitly reflected in the selection of index. As a result, the theoretical received power scaling law is summarized in $\emph{Proposition 1}$.

$\emph{Proposition 1}$: Assume $h_{r,n}$ following the Rician channel model with Rician factor of $F_r$ , $n = 1, \cdots, N$. For $N \gg 1$, the average signal power received at the user is given by
\begin{align}\label{eq:25}
& P_r = P_d M \beta_r \beta_g \left( N^2 F_1^2 + N F_1 F_2 \sqrt{\pi} \ + \right. \notag \\ & \left. N F_2^2 \frac{N + \frac{\pi}{2} (N-1) \sqrt{\frac{\beta_r \beta_g}{(N-1)\beta_r\beta_g + \sigma_q^2}} }{N+\frac{\pi}{2} \sqrt{N-1}} \left( \text{log}Q + C \right) \right) ,
\end{align} where $F_1 =  \sqrt{F_r/(F_r+1)}$, $F_2 = \sqrt{1/(F_r+1)}$ and $C\approx0.57722$ is the Euler-Mascheroni constant.

$\emph{Proof}$: Adopting the above conditions, the signal power received at the user can be expressed as
\begin{align} \label{eq:Pr}
    P_r & = P_d \mathbb{E} \left\{ \max_{q \in \mathcal{Q}} \left\lVert \sum_{n=1}^{N} h_{r,n}^* \varphi_{q,n} \mathbf{g}_n^T \right\rVert^2 \right\} \notag \\ & \le P_d \beta_r \beta_g M \notag\\ & \times \mathbb{E} \left\{\max_{q \in \mathcal{Q}} \left|\sum_{n=1}^{N} \left( F_1 h_{r,n}^{*\text{LoS}}+F_2 h_{r,n}^{*\text{NLoS}} \right) \varphi_{q,n} g_{n,m}^{\text{LoS}} \right|^2 \right\} \notag \\ & = P_d \beta_r \beta_g \mathbb{E} \left\{ \max_{q \in \mathcal{Q}} \left\{ \left|\sum_{n=1}^{N} F_1 h_{r,n}^{*\text{LoS}} \varphi_{q,n} g_{n,m}^{\text{LoS}}\right|^2 \right. \right. \notag\\ & \left. \left. +  \left|\sum_{n=1}^{N} F_2 h_{r,n}^{*\text{NLoS}} \varphi_{q,n} g_{n,m}^{\text{LoS}}\right|^2 + 2 \Re \left \{ \left(\sum_{n=1}^{N} h_{r,n}^{*\text{LoS}} \varphi_{q,n} g_{n,m}^{\text{LoS}}\right) \right. \right. \right. \notag \\ & \left. \left. \left. \times \left(\sum_{n=1}^{N} h_{r,n}^{*\text{NLoS}} \varphi_{q,n} g_{n,m}^{\text{LoS}}\right)^* F_1 F_2  \right \} \right\} \right\}, \end{align} 
where $\mathbf{g}_n^T$ is the $n$th row of $\mathbf{G}$ and ‘$\le$’ reduces to ‘=’ if and only if $\text{rank}(\mathbf{G}) = 1$. As we generate virtual channels based on statistical CSI, the LoS component of the channel is always aligned, then we can obtain the result $N^2 F_1^2$ by applying (31) in \cite{r19}. Next, since we have $\sum_{n=1}^{N} \varphi_{q,n} h_{r,n}^{*\text{NLoS}} g_{n,m}^{\text{LoS}} \sim \mathcal{CN} \left(0, N\right)$ as $N \to \infty$ based on Lindeberg-Levy central limit theorem \cite{r19}, the third entry of (\ref{eq:Pr}) can be further written as $2 F_1 F_2 \Re \left\{ \mathbb{E} \left\{ \max_{q \in \mathcal{Q}} \left\{ \sum_{n=1}^{N} h_{r,n}^{*\text{LoS}} h_{r,n}^{\text{NLoS}} \right\} \right\} \right\}$, which has a upper bound of $2 F_1 F_2 \sum_{n=1}^{N} \left( \left|h_{r,n}^{\text{LoS}}\right| \mathbb{E} \left\{ \left|h_{r,n}^{*\text{NLoS}}\right| \right\} \right) = \sqrt{\pi} F_1 F_2 N$. Furthermore, the second entry reflects the effects on the randomly generated NLoS component. By applying (24) in \cite{r15}, we can obtain the result $N(\text{log}Q+C)$. Moreover, we multiply the result of the perfect CSI scenario by a factor $\mu$ to get an upper bound. Specifically, let $h_{c,n}^* = \sqrt{\beta_r} h_{r,n}^{*\text{NLoS}} \sqrt{\beta_g}g_{n,m}^{\text{LoS}}$ represent the NLoS component of the cascaded channel spanning from the user to BS through the $n$th RIS element, according to the achievable rate maximization problem, we have the optimal RC with perfect and imperfect CSI as $\bar{\varphi}_{q,n} = \frac{h_{c,n}}{\left| h_{c,n}^* \right|} \frac{\sum_{i \neq n}^{N} h_{c,i}^* \varphi_{q,i} }{\left| \sum_{i \neq n}^{N} h_{c,i}^* \varphi_{q,i} \right|}$ and $\hat{\varphi}_{q,n} = \frac{h_{c,n}}{\left|h_{c,n}^* \right|} \frac{\sum_{i \neq n}^{N} h_{c,i}^* \varphi_{q,i} + \varepsilon_q}{\left| \sum_{i \neq n}^{N} h_{c,i}^* \varphi_{q,i} + \varepsilon_q \right|}$, respectively. Substituting them into the second entry of (\ref{eq:Pr}), we have $\mu = \frac{N + \frac{\pi}{2} (N-1) \sqrt{\frac{\beta_r \beta_g}{(N-1)\beta_r\beta_g + \sigma_q^2}} }{N+\frac{\pi}{2} \sqrt{N-1}}$. Thus, we complete the proof. $\hfill\blacksquare$

It is noted that when the channel estimation errors are adequately small, the theoretical received power in the perfect CSI scenario can be asymptotically obtained by $P_r = P_d \beta_r \beta_g M \left(F_1^2 N^2 + F_2^2 N \left(\text{log}Q+C\right) + \sqrt{\pi} F_1 F_2 N \right)$.

\section{Simulation Results}

In this section, we provide simulation results to validate the performance of the proposed scheme. All results are obtained by averaging 1,000 independent experiments. We consider a 3D Cartesian coordinate system, where the antenna array at the BS is modeled by a uniform linear array and deployed at a height of $h_{\text{BS}} = \SI{5}{m}$ on the $z$-axis with antenna spacing of $d_{\text{BS}} = \lambda/2$. The number of the BS antennas is set to $M = 8$. The RIS is modeled by a uniform planar array and deployed at a height of $h_{\text{R}} = \SI{5}{m}$ on the $y\text{-}z$ plane with $10 \times 10$ array structure ($N = 100$) with element spacing of $d_{\text{R}} = \lambda/8$. The distance between BS and RIS is $d_{\text{BR}} = \SI{100}{m}$. Let $d_0 = \SI{2}{m}$, we assume two users located at $(d_0, d_{\text{BR}}, 0)$ and $(-d_0, d_{\text{BR}}, 0)$, respectively. The Rician factor of $\mathbf{G}$, $\mathbf{h}_r$ and $\mathbf{h}_d$ are $F_g = \SI{4}{dB}$, $F_r = \SI{3}{dB}$ and $F_d = \SI{-3}{dB}$, respectively. The path loss factor of $\mathbf{G}$, $\mathbf{h}_r$ and $\mathbf{h}_d$ are $\alpha_g = 2.4$, $\alpha_r = 2.5$ and $\alpha_d = 3.5$, respectively. The quantization number is set to $b = 1$. The path loss is modeled as $\beta = C_0(d/d_0)^{-\alpha}$, where $d$ is the distance of corresponding link, $C_0 = \SI{-20}{dB}$ denotes the path loss of the reference distance $d_0 = \SI{1}{m}$ and $\alpha$ denotes the path loss factor. Moreover, the transmit power at the BS is set to $P_d = \SI{40}{dBm}$, the average noise power at the BS and the user are $\sigma_z^2 = \SI{-110}{dBm}$ and $\sigma_k^2 = \SI{-90}{dBm}$, respectively. And the average power of the pilot signals is $P_{\text{ul}} = \SI{-20}{dBm}$.

\begin{figure}[!t]    
\centering            
\subfloat[\scriptsize Perfect CSI]  
{
\label{fig_3:subfig1}\includegraphics[width=0.225\textwidth]{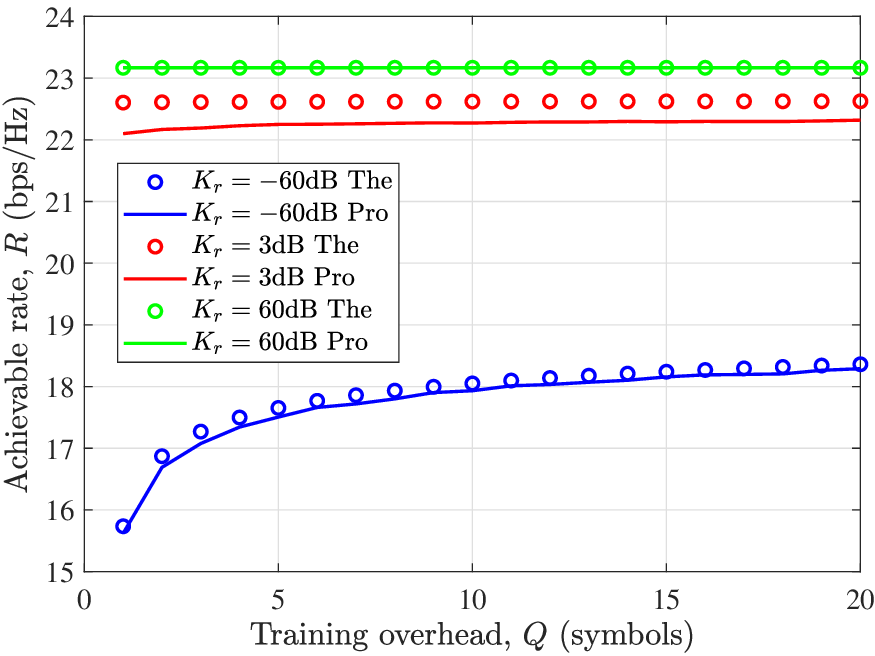}
}
\subfloat[\scriptsize Imperfect CSI]
{
\label{fig_3:subfig2}\includegraphics[width=0.225\textwidth]{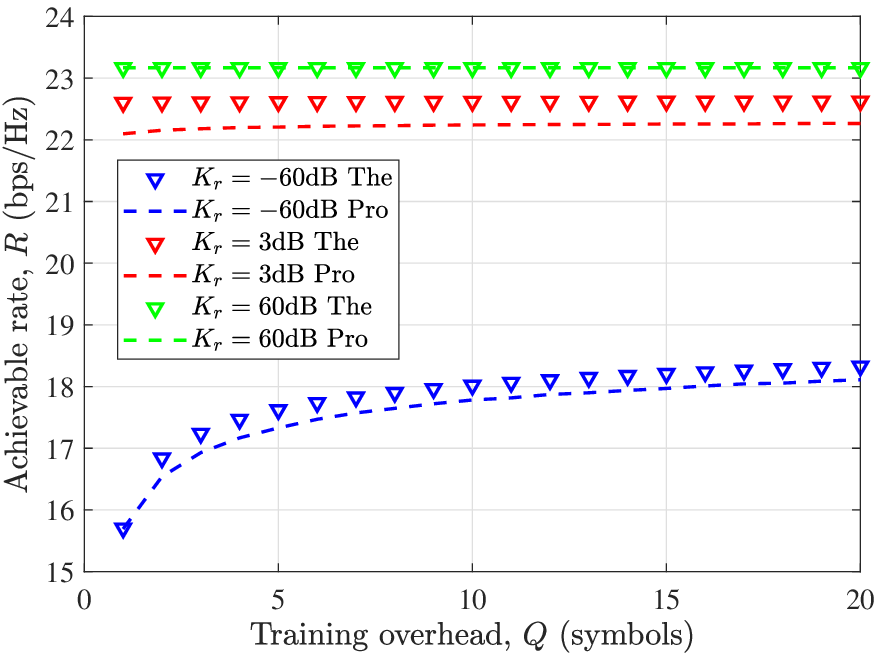}
}

\caption{ Achievable rate, $R$ versus the training overhead, $Q$.}   
\label{fig_3}       
\end{figure}

Firstly, we verify the performance analysis in a single user scenario, where we consider a user located at $(d_0, d_{\text{BR}}, 0)$. As shown in Fig. \ref{fig_3}, we verify our analytical results under different RIS-user channel conditions, where we set $F_r = \SI{-60}{dB}$, $\SI{3}{dB}$, and $\SI{60}{dB}$, respectively. It can be seen from Fig. \ref{fig_3} that the proposed scheme can gradually improve the achievable rate with the increase of the training overhead. Moreover, in both the scenario of perfect CSI and imperfect CSI, the derived theoretical performance bound serves a tight upper bound of the simulation results. 

\begin{figure}[!t]    
\centering     

\subfloat[\scriptsize]  
{
\label{fig_5:subfig1}\includegraphics[width=0.225\textwidth]{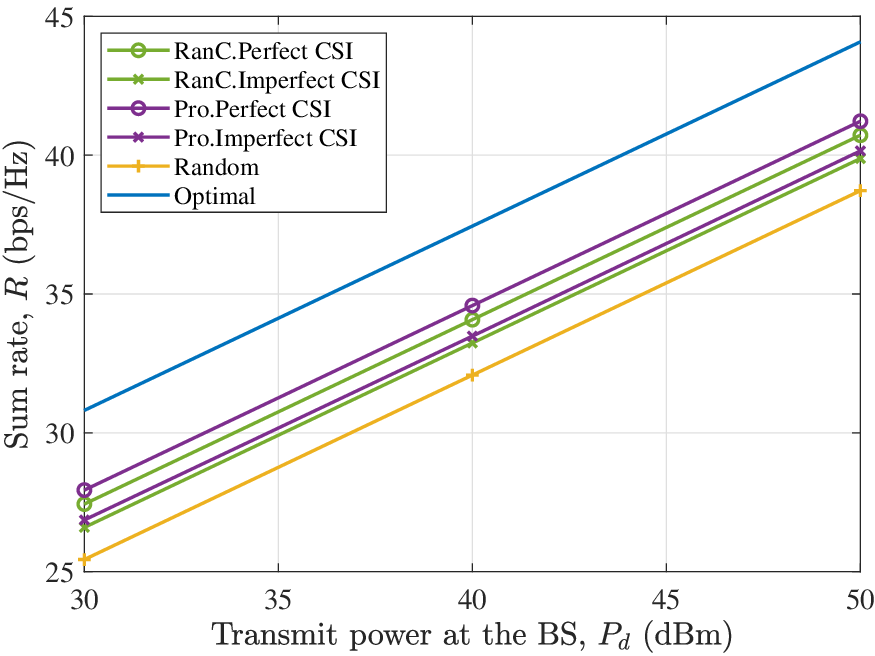}
}
\subfloat[\scriptsize]
{
\label{fig_5:subfig2}\includegraphics[width=0.225\textwidth]{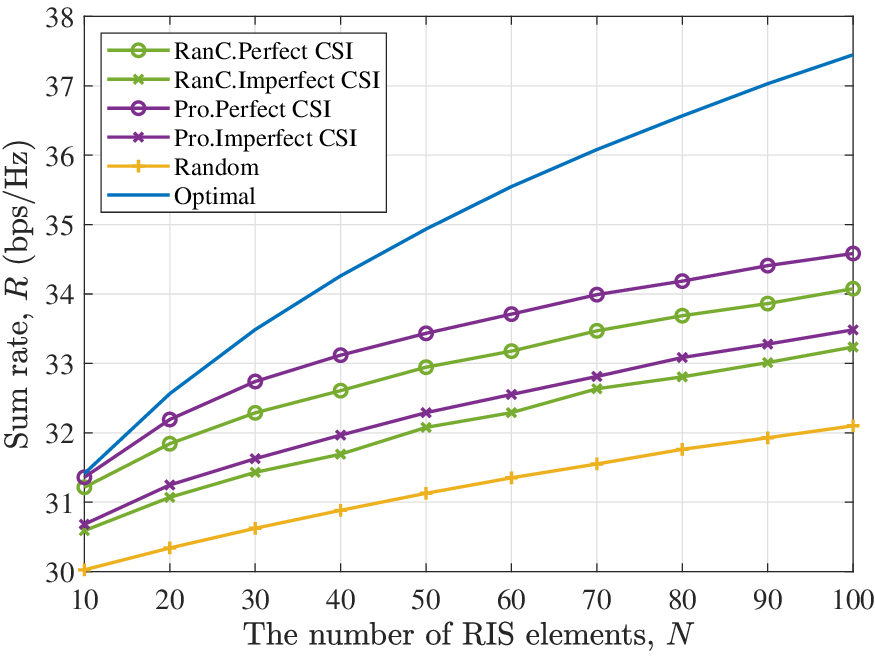}
}
        
\subfloat[\scriptsize]  
{
\label{fig_5:subfig3}\includegraphics[width=0.225\textwidth]{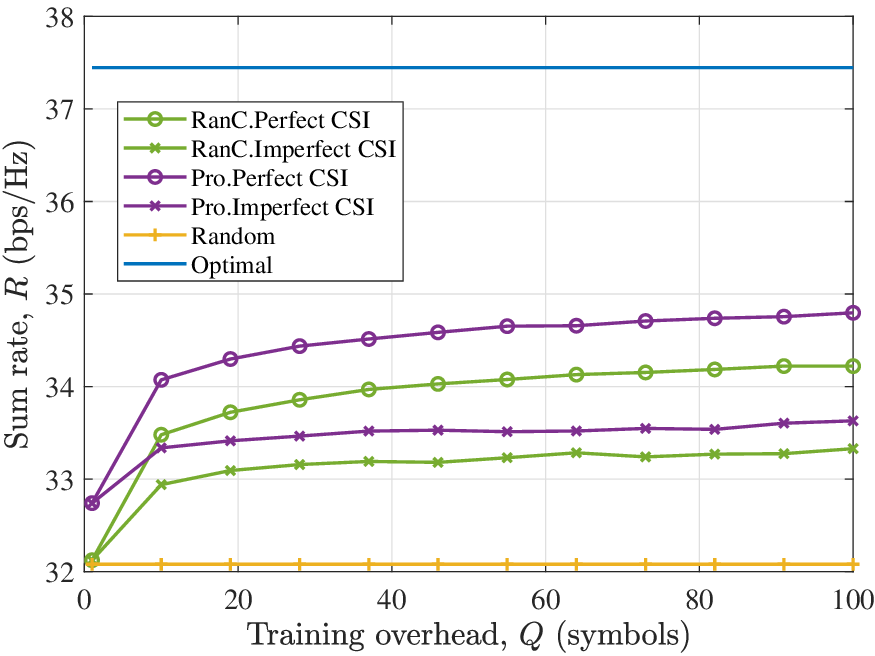}
}
\subfloat[\scriptsize]
{
\label{fig_5:subfig4}\includegraphics[width=0.225\textwidth]{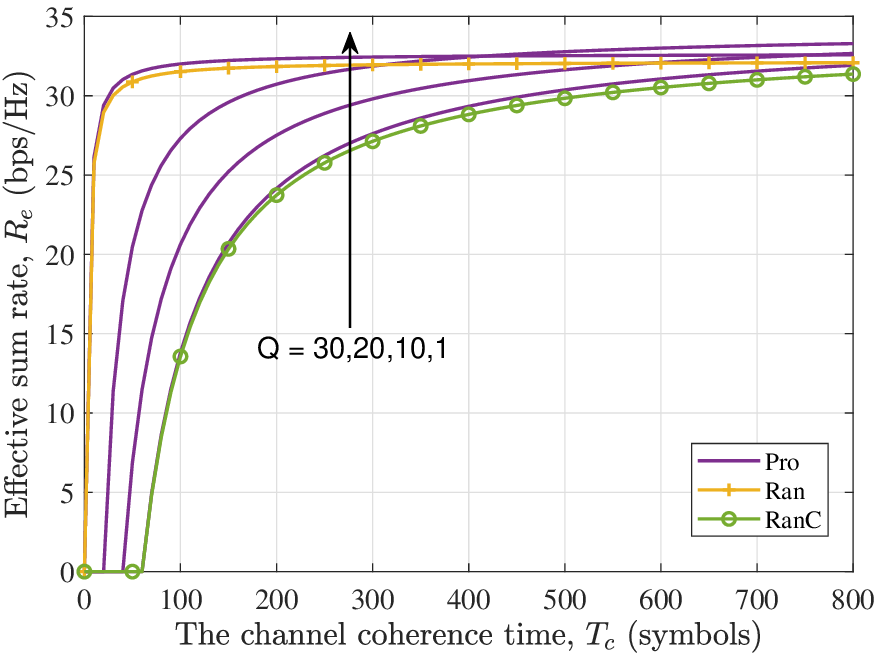}
}
\caption{(a) Sum rate, $R$ versus the transmit power at BS, $P_d$. (b) Sum rate, $R$ versus the number of RIS elements, $N$. (c) Sum rate, $R$ versus the training overhead, $Q$. (d) Effective sum rate, $R_e$ versus the channel coherence time, $T_c$ with perfect CSI.}   

\label{fig_5}   
\end{figure}

Fig. \ref{fig_5:subfig1} evaluates the sum rate versus the total transmit power, where the codebook size is set to $Q = 50$. We can observe a logarithmic increase relationship between the sum rate and the transmit power at the BS, which corresponds to our theoretical formula. In addition, the proposed scheme outperforms the random codebook scheme in both CSI scenarios. This is because the proposed scheme optimizes the RIS RCs by leveraging the statistical CSI. Then, as shown in Fig. \ref{fig_5:subfig2}, with the increase of the RIS elements $N$, the sum rate of the proposed scheme with diminishing return, which is due to the fact that a larger number of reflecting elements could reap more radiating energy by coherently superimposing all cascaded links. 

Next, Fig. \ref{fig_5:subfig3} examines the effects of the overhead on the performance of the proposed scheme. It is noted that as the training overhead $Q$ increases, the proposed scheme gets a certain promotion in terms of the sum rate and always outperforms the random codebook scheme as it utilizes statistical CSI to design the RC. Finally, Fig. \ref{fig_5:subfig4} examines the effective sum rate versus the channel coherence time. Specifically, the effective achievable rate is defined as $R_e = \frac{T_c-\tau}{T_c} \mathbb{E} \left\{ \sum_{k=1}^{K} \text{log}_2 \left( 1 + \frac{\left| \mathbf{h}_{k}^H \mathbf{w}_{k} \right|^2}{\sum_{l \neq k}^K \left| \mathbf{h}_{k}^H \mathbf{w}_{l} \right|^2 + \sigma_k^2} \right) \right\}$, where $T_c$ is the channel coherence time and $\tau$ is the pilot overhead. We can observe in Fig. \ref{fig_5:subfig4} that the proposed scheme can flexibly adjust the training overhead according to the specific channel coherence time to achieve the highest effective sum rate.

\section{Conclusion}

In this paper, we proposed a channel training-based protocol and an environment-aware codebook generation scheme for RIS-assisted MU-MISO systems. Specifically, the proposed scheme generated a set of virtual channels based on statistical CSI and obtained the codewords according to the AO method offline. During the on-line stage, we configured the RIS according to the pre-designed codewords and performed the composite channel estimation and transmit precoding design. Finally, the optimal codeword resulting in the highest sum rate was adopted to configure the RIS for assisting in the data transmission. Furthermore, we analyzed the performance of the proposed scheme accounting for the imperfect CSI scenario and evaluated the achievable rate of the environment-aware codebook scheme through simulation results.

\bibliographystyle{IEEEtran}
\bibliography{IEEEabrv,Ref}

\end{document}